\documentclass[pre,aps,twocolumn,showpacs,superscriptaddress]{revtex4}

\def \beq{\begin{equation}}         \def \eeq{\end{equation}}
\def \beqa{\begin{eqnarray}}        \def \eeqa{\end{eqnarray}}
\def \bea{\begin{array}}        \def \eea{\end{array}}

\def\jmb#1#2#3{{J. Mol. Biol. }{\bf #1}, #2 (#3)}

\def\natrg#1#2#3{{Nature Rev. Genet. }{\bf #1}, #2 (#3)}
\def\natg#1#2#3{{Nature Genet. }{\bf #1}, #2 (#3)}
\def\pnas#1#2#3{{Proc. Natl. Acad. Sci. USA }{\bf #1}, #2 (#3)}

\def\prl#1#2#3{{Phys. Rev. Lett. }{\bf #1}, #2 (#3)}
\def\sci#1#2#3{{Science }{\bf #1}, #2 (#3)}

\usepackage{amsmath}
\usepackage{epsfig}
\usepackage[dvips]{color}
\begin{document}

\title{Response delay as a strategy for survival in fluctuating
environment}
\author{Xiao chuan Xue}
\affiliation{Center for Advanced Study, Tsinghua University,
Beijing, 100084, China}
\author{Jinhua Zhao}
\affiliation{Institute of Theoretical
Physics, The Chinese Academy of Sciences, P.O.Box 2735 Beijing
100080, China}
\author{Fei Liu}
\email[Email address:]{liufei@tsinghua.edu.cn}
\affiliation{Center for Advanced Study, Tsinghua University,
Beijing, 100084, China}
\author{Zhong-can Ou-Yang}
\affiliation{Center for Advanced Study, Tsinghua University,
Beijing, 100084, China} \affiliation{Institute of Theoretical
Physics, The Chinese Academy of Sciences, P.O.Box 2735 Beijing
100080, China}
\date{\today}

\begin{abstract}
{Response time-delay is an ubiquitous phenomenon in biological
systems. Here we use a simple stochastic population model with
time-delayed switching-rate conversion to quantitatively study the
biological influence of the response time-delay on the survival
fitness of cells living in periodically fluctuating and stochastic
environment, respectively. Our calculation and simulation show
that for the cells having a slow rate transition into fit
phenotype and a fast rate transition into unfit phenotype, the
response time-delay can always enhance their fitness during the
environment change. Particularly, in the periodic or stochastic
environment with small variance, the optimal fitness achieved by
these cells is superior to that of cells with reverse switching
rates even if the latter exhibits rapid response. These results
suggest that the response time delay may be utilized by cells to
enhance their adaptation to the fluctuating environment. }
\end{abstract}
 \maketitle

\section{Introduction}
Most of living species are exposed to random fluctuations in their
environment in the Nature. Environment change can either be
periodic, such as day-night interchanges, or be stochastic, such as
nutrition supplements in random times. Thus how can a life-form
adapt to the fluctuating environment to optimize its fitness within
is critical for the species to survive in the
nature~\cite{Schaffer,Mroz}. Recently, with the progresses in the
study of the gene regulation networks, the stochastic nature of the
gene expression is revealed~\cite{Samoilov,Kaern,Kaufmann}, which is
soon considered to further understand optimized fitness of cells.
Stochastic gene expression brings the isogenic population to
distinct phenotypes which is known as phenotypic
heterogeneity~\cite{Samoilov,Lu}. Phenotypic heterogeneity is a
useful strategy for species to adapt to the fluctuating
environment~\cite{Sasaki}. With this concept, various phenomena,
including the bacterial persistence~\cite{Balaban,Bishop,Kussell}
and optimized growth rate for the cells living in the fluctuating
environment~\cite{Thattai,Kussell1,Acar,Pang,Kussell2}, were
studied. These authors also considered  comparison between the
strategy of the heterogeneity in which the switching rate from the
fit state to the unfit states in an environment of specific
parameters is nonzero and the homogeneity in which the switching
rate mentioned above is zero~\cite{Thattai,Pang}, the effects of the
static (switching rates do not change with environment) and dynamic
(switching rates change with environment) transition between
states~\cite{Kussell1}, and the influence of the switching rates on
the environment changing rate~\cite{Acar}, which are all important
for the understanding of the survival of species on earth.

Besides the stochastic gene expression, another important feature of
gene regulation networks is time-delay. A time delay usually results
from regulated transcription and translation, and leads to behaviors
quite different from those without~\cite{Bratsun}. In more complex
gene networks, some motifs, such as feed-forward loops and gene
cascades, induce time-delays of length even longer than a cell
cycle~\cite{Rosenfeld,Mangan,Mangan1}. These time-delays have
important effects on gene regulatory networks, and there were some
assumption that cells can benefit from their time-delay motif in
stochastically fluctuating environments~\cite{Dekel}. However, so
far as we know, there is yet no work on the influence of the
response time-delay to the fitness of cells living in the
fluctuating environment. In this paper, we construct a stochastic
population model with time-delayed
 switching-rate conversation to quantify the biological
influence of the response time-delay on the survival of cells in
fluctuating environments.

\section{Model}
To study the fitness of cell population, we present a stochastic
population model~\cite{Thattai}, in which environment parameters
fluctuate between two different states ${\tt E}_{\tt a}$ and ${\tt
E}_{\tt b}$, and accordingly in which cells display two different
phenotypes ${\tt a}$ and ${\tt b}$. Cells of phenotype ${\tt a}$ are
fit for growth in the environment ${\tt E}_{\tt a}$ with fast growth
rate $\gamma_{\tt f}$, while those of ${\tt b}$ is unfit for growth
in ${\tt E}_{\tt a}$ with slow $\gamma_{\tt u}$. Situation reverses
in the environment ${\tt E}_{\tt b}$ as shown in figure 1(a). The
environment compensation for the growth of the fit cells is $\Delta
\gamma=\gamma_{\tt f}-\gamma_{\tt u} > 0$. In addition, in these two
environment states, cells stochastically switch between the two
distinct phenotypes with switching rate $k_{\tt f}$ into fit
phenotype and $k_{\tt u}$ into unfit one. Some authors generally
treated the switching-rate conversion simultaneous with the
environment change~\cite{Thattai,Pang}. But here in order to
consider the influence of response time-delay, we set the switching
rates conversion with a delayed time $\tau$ after an state change;
see figure 1(a). The delayed time $\tau$ is a consequence of the
gene networks structure when responding to environment changes. In
this model, the dynamics of the cell numbers in the fit ($n_{\tt
f}$) and unfit ($n_{\tt u}$) states can be described as
\begin{eqnarray}
&&\frac{d}{dt}n_{\tt f}=\gamma_{\tt f}n_{\tt f}+k_{\tt u}n_{\tt
u}-k_{\tt f}n_{\tt f},\;0<t<\tau;\nonumber\\
&&\frac{d}{dt}n_{\tt u}=\gamma_{\tt u}n_{\tt u}+k_{\tt f}n_{\tt
f}-k_{\tt u}n_{\tt u},\;0<t<\tau;\nonumber\\
&&\frac{d}{dt}n_{\tt f}=\gamma_{\tt f}n_{\tt f}+k_{\tt f}n_{\tt
u}-k_{\tt u}n_{\tt f},\;\tau<t<T;\nonumber\\
&&\frac{d}{dt}n_{\tt u}=\gamma_{\tt u}n_{\tt u}+k_{\tt u}n_{\tt
f}-k_{\tt f}n_{\tt u},\;\tau<t<T,\label{number}
\end{eqnarray}
where $T$ is the time cells dwell in an environment state before
change to the other state. We assume $\tau<T$, which means that
dynamic transition condition holds during the switching-rates
conversion. When $\tau>T$, switching rates do not convert in this
environment state which in fact becomes the static transition.
However, static transition can also be a possible choice for cells,
and below we consider above situation together.

To investigate the fitness of cells, we consider the time-averaged
population growth-rate $\langle \gamma \rangle=\langle \gamma_{\tt
u} f_{\tt u}(t)+\gamma_{\tt f} f_{\tt f}(t) \rangle=\gamma_{\tt
u}+\Delta\gamma \langle f_{\tt f}(t)\rangle$, where $f_{\tt
i}(t)={n_{\tt i}}/{(n_{\tt u}+n_{\tt f})},\ {\tt i=u,f}$ is the
fraction of cells in the unfit and fit phenotypes in the two
environment states respectively. We assume in this model that the
fractions of the cells in fit and unfit phenotypes convert when the
cells respond to environment changes, that's, if the fraction of the
cells in the fit phenotype is $f_{\tt f}$ just before the
environment turnover, it will be $1-f_{\tt f}$ after. Hereafter we
focus on the time evolution of the $f_{\tt f}$ and its time-average
($\langle f_{\tt f}\rangle$) which accounts for the fitness of the
cell population. By applying equation~\ref{number}, the dynamics of
$f_{\tt f}$ can be described by the following equations:
\begin{eqnarray}
\frac{d}{dt}f_{\tt f}(t)=k_{\tt u}+(\Delta\gamma-k_{\tt u}-k_{\tt
f})f_{\tt f}-\Delta\gamma
f_{\tt f}^2,\;0<t<\tau;\nonumber\\
\frac{d}{dt}f_{\tt f}(t)=k_{\tt f}+(\Delta\gamma-k_{\tt u}-k_{\tt
f})f_{\tt f}-\Delta\gamma f_{\tt f}^2,\;\tau<t<T,\label{fraction}
\end{eqnarray}
In different fluctuating environment conditions, $T$ and $\tau$ can
be tuned. we solve equations~\ref{fraction}, then average $f_{\tt
f}$ with time to evaluate the influence of response time delay on
the survival fitness of cells.

\begin{figure}[htpb]
\begin{center}
\includegraphics[width=1.\columnwidth]{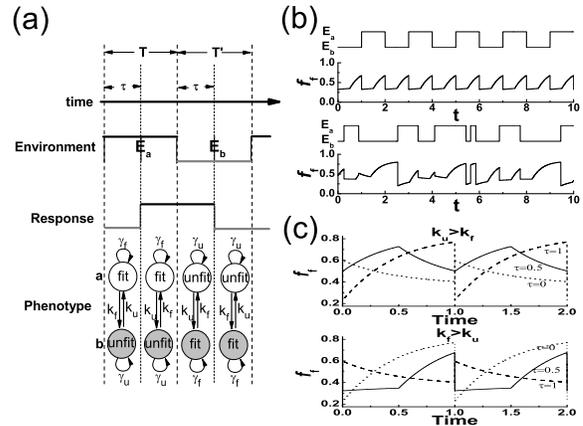}
\caption{(a) The scheme of the response time delay model of the
fluctuating environment. (b) $f_(\tt f)$ vs time in the periodic and
stochastic environments, respectively. Corresponding fraction of the
fit cells ($f_{\tt f}$) is plotted, while parameters are set as
$k_{\tt u}=0.5,\ k_{\tt f}=1.5,\ \Delta\gamma=1$ and $\tau=0.5$. (c)
Growth in a symmetrically periodic environment. The fraction of the
fit cells ($f_{\tt f}$) is plotted with time for a period. In the
figure we show the results of three specific $\tau$, 0, 0.5 and 1.
The parameters in the upper and lower plots are set as $k_{\tt
u}=1.5,\ k_{\tt f}=0.5,\ \Delta\gamma=1$ and $k_{\tt u}=0.5,\ k_{\tt
f}=1.5,\ \Delta\gamma=1$, respectively.} \label{figure1}
\end{center}
\end{figure}

\section{Result and Discussion}
Figure~\ref{figure1} (b) shows two examples of $f_{\tt f}(t)$
changing  with time in periodic and stochastic environments. In the
periodic environment state, cells spends a specific time $T$ in each
states, while in the stochastic case, the time spent in each states
has a mean value $T$. Without losing generality, we set $T=1$. From
the figure we can see that after considering the time delay, in the
periodic environment there will be an extra turning point at time
$\tau$ after each environment change, while in the stochastic
environment, only the environment states with duration time longer
than $\tau$ show a state turning point. Those brief environment
states with duration time shorter than $\tau$ are filtered as a
result of phenotype switching system of cells. The additional
turning points of the $f_{\tt f}(t)$ in different environments and
filter out of the short duration environment states in the
stochastic environment may introduce intriguing behavior in
population fitness of cells with response time-delay, as we show
below.

To investigate the influence of response time-delay, we first
consider a simple symmetrically periodic environment with a specific
time-delay~\cite{xue}. For different $\tau$, $f_{\tt f}(t)$ reaches
its periodic state as shown in figure 1(c). Two cases are shown
there with inter-conversed switching-rates. For $\tau=0$, the model
returns to what is studied by Thattai {\it et. al}, and so does when
$\tau=1$, in which $k_{\tt u}$ denotes the rate switching to the fit
state and $k_{\tt f}$ to the unfit one. Therefore, from
figure~\ref{figure1} (c), we can see that the pattern of $f_{\tt
f}(t)$ in the upper graph when $\tau=0$ is the same as the lower one
for $\tau=1$, and so does when $\tau=1$ and $\tau=0$ . Since
$\langle f_{\tt f}(t)\rangle$ reflects the fitness, we then
calculate its averaged value. For these two delayed time discussed
above, one can easily notice that the case with larger switching
rate to the fit phenotype generally have higher $\langle f_{\tt
f}\rangle$; see figure 2(a). When $0<\tau<1$, the tuning point in
$f_{\tt f}(t)$ (figure~\ref{figure1} (b), (c)) makes $\langle f_{\tt
f}\rangle$ behaviors quite different from those without time delays.
After some mathematical analysis and with the help of numerical
solution, we find that $\langle f_{\tt f}\rangle$ both decrease
where $\tau\rightarrow 0^+$ and $1^-$, indicating a maximum at an
intermediate $\tau$ when $k_{\tt u}>k_{\tt f}$. While $k_{\tt
f}>k_{\tt u}$, fitness increases both as $\tau\rightarrow 0^+$ and
$1^-$, indicating a minimum at an intermediate $\tau$ (figure 2).
Besides, there is another case when $\tau>1$, when switching rates
have no enough time to response to the environment change in neither
environment states, which is just the case of static transition. In
this case, the switching rate to the fit phenotype is half $k_{\tt
f}$ and half $k_{\tt u}$ in a whole period, and the average effect
should be some combination of the $k_{\tt u}$ and $k_{\tt f}$. So
the value of $\langle f_{\tt f}\rangle$ is between the two cases of
$\tau=0$ and $\tau=1$ (figure 2(a)).

\begin{figure}
\begin{center}
\includegraphics[width=1.\columnwidth]{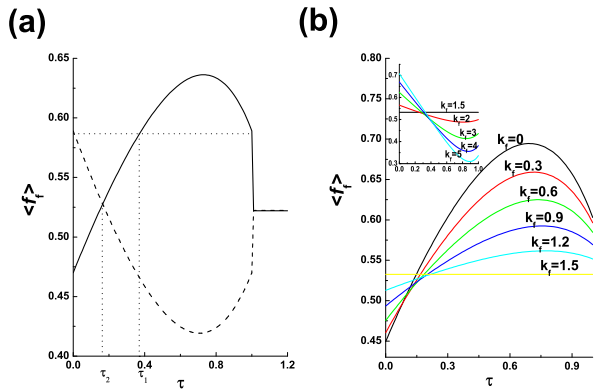}
\caption{(Color online). Influence of the response time-delay on the
time-averaged fraction of cells in fit phenotype. (a) Time-averaged
$f_{\tt f}$ is plotted with $\tau$. The solid and dash lines denote
the case when $k_{\tt u}=1.5>k_{\tt f}=0.5$ and $k_{\tt
f}=1.5>k_{\tt u}=0.5$, respectively, as $\Delta \gamma=1$. Two cases
of specific time-delays are also shown in the figure. (b) Parameter
influence on the time-averaged $f_{\tt f}$. Here we set $k_{\tt
u}=1.5$, $\Delta\gamma=1$, and $k_{\tt f}\leq k_{\tt u}$ from 0 to
1.5. Insert shows $k_{\tt f}\geq k_{\tt u}$ from 1.5 to 5. }
\label{figure2}
\end{center}
\end{figure}

The results shown above correspond to certain sets of parameters. In
all there are three parameters that can be tuned in this model.
$\Delta \gamma$ represents the growth compensation in different
environment. Large $\Delta \gamma$ indicates much faster cells
growth in fit environment than in unfit one, thus it has significant
effects on $\langle f_{\tt f}\rangle$, but limited effects on the
trend of $\langle f_{\tt f}\rangle$ to the time-delay $\tau$. Larger
$\Delta \gamma$ induces higher $\langle f_{\tt f}\rangle$, but make
smaller the difference of $\langle f_{\tt f}\rangle$ induced by
different $\tau$. However, an extremum at certain $\tau$ always
exist(data not shown). The other two tunable parameters are $k_{\tt
u}$ and $k_{\tt f}$. They show similar effects in our model(equation
1). Here we first keep $k_{\tt u}$ fixed and vary $k_{\tt f}$ to see
its influence on the $\langle f_{\tt f}\rangle$. As shown in figure
2(b), when $k_{\tt f}=k_{\tt u}$, exchange of switching rates become
trivial in the model, thus delay times have no influence on $\langle
f_{\tt f}\rangle$ and $\langle f_{\tt f}\rangle$ remains constant at
different $\tau$. When $k_{\tt f}<k_{\tt u}$, there is a maximum
$\langle f_{\tt f}\rangle$ at a certain $\tau$, even in the
homogenetic case when ($k_{\tt f}=0$). In addition, in the delayed
time (0$\sim\tau$ with $k_{\tt u}$), larger ${k_{\tt u}}/{k_{\tt
f}}$ indicates a faster transition into the fit phenotype, thus
results in a steeper slope of the $\langle f_{\tt f}\rangle$ while
reaching its maximum, which means more significant influence of
response time-delay on increasing the fitness in case of larger
${k_{\tt u}}/{k_{\tt f}}$ (figure 2(b)). A contrary effect is
observed for $k_{\tt f}>k_{\tt u}$, which always shows a
minimum(figure 2(b) insert). The slope of the decreasing $\langle
f_{\tt f}\rangle$ is steeper with larger ${k_{\tt f}}/{k_{\tt u}}$,
which means the response time-delays also show more significant
negative influence on the fitness with larger ${k_{\tt f}}/{k_{\tt
u}}$ (figure 2(b) insert). These results verify that the influence
of the response time delay on the fitness of cells is universal and
does not strictly depend on state parameters.

These results show some interesting behavior of the response time
delay on the fitness of cells. First, if $k_{\tt f}>k_{\tt u}$ which
means a larger transition rate into fit phenotype in each
environment, the best strategy to enhance fitness is to reduce the
response time delay, as indicated in figure~\ref{figure2} (b) and
larger ${k_{\tt f}}/{k_{\tt u}}$ introduces a more severe defect in
fitness. Second, above results tells us that cells do not always
need to adopt $k_{\tt f}>k_{\tt u}$ to achieve a higher fitness. If
cells use some special strategy to reverse the value of the two
switching rates to set $k_{\tt u}>k_{\tt f}$, the response time
delay can always approve cells' fitness. Besides, there is a
specific delay time $\tau_1$ after which $\langle f_{\tt f}\rangle$
is larger than what cells can reach without time delay in the
similar set of parameters; see figure 2(a). In the region
$\tau_1<t<1$, the response time delays prove to be an effective way
for cells of optimizing their fitness. In addition, considering the
constant presence of an intrinsic response time-delay in gene
networks, there are two switching rates cells can choose. For
example, when the delay time is longer than $\tau_2$, a strategy
which sets $k_{\tt u}$ to be the larger one gets a better fitness
(figure 2(a)). The parameter-dependence of $\tau_1$ and $\tau_2$ are
shown in figure 3(a). The value of $\tau_1$ is always smaller than
1, which confirms the independence of parameters of these effects,
relatively. Figure 3(a) also shows a large parameter space to be
realized to optimize fitness by response time-delay. Third, the
static transition strategy ($\tau>1$) can be a choice for cells in
some special cases. For example, when the systemic time delay is
large than a certain value with $k_{\tt f}>k_{\tt u}$ or shorter
with $k_{\tt f}<k_{\tt u}$ (figure 2(a)), static transition is
better than dynamic one in enhancing survival fitness, while in
other conditions with the same switching rates, dynamic transition
strategy is still an better option. Finally, following the finding
that $f_{\tt f}(t)$ for $\tau=0$ and 1 are the same while exchanging
$k_{\tt f}$ and $k_{\tt u}$, we find that after a similar
manipulation of $\tau$ with $1-\tau$ and $k_{\tt f}$ with $k_{\tt
u}$ in equation 1, this new equation becomes one describing a system
with a pre-response time $\tau$. Therefore, the $\langle f_{\tt
f}\rangle$ dependence on the pre-response time $\tau$ in the new
equation for $k_{\tt f}>k_{\tt u}$ is in fact the same as those with
time-delay $\tau$ in the former equation with $k_{\tt f}<k_{\tt u}$,
which means that an advance-response may induce a better fitness for
the population when the cells adopt a faster switching rate to the
fit phenotype~\cite{Baliga,Tagkopoulos}.

\begin{figure}
\begin{center}
\includegraphics[width=1.\columnwidth]{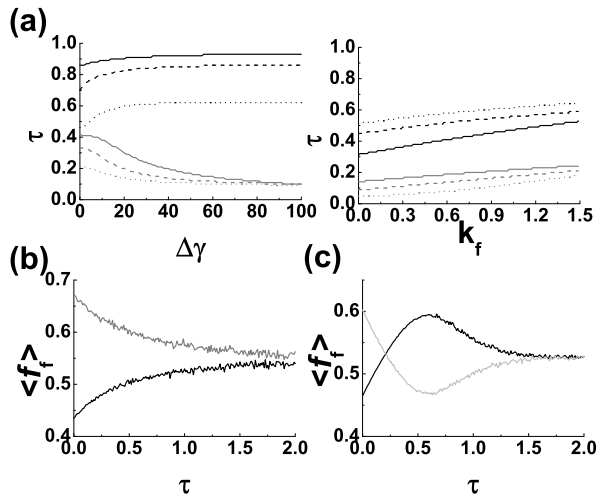}
\caption{(a)Parameter-dependence of the two special delay times.
Black and grey lines denote $\tau_1$ and $\tau_2$, respectively. On
the left is their dependence on $\Delta\gamma$, with $k_{\tt f}=10,\
k_{\tt u}=0$ (solid line), $k_{\tt f}=5,\ k_{\tt u}=0.1$ (dashed
line), and $k_{\tt f}=1.5,\ k_{\tt u}=1$ (dotted line). On the right
shows their dependence on $k_{\tt f}$, with $k_{\tt u}=1.5$ and
$\Delta\gamma=$ 1 (solid line), 10 (dashed line), and 30 (dotted
line), respectively. (b) and (c) Cell growth in the fluctuating
environment. Time-averaged $f_{\tt f}$ is plotted with $\tau$, with
$n=1$ in (b) and $n=10$ in (c). The black and grey lines denote the
case when $k_{\tt u}=1.5>k_{\tt f}=0.5$ and $k_{\tt f}=1.5>k_{\tt
u}=0.5$, respectively. $\Delta\gamma=1$.} \label{figure3}
\end{center}
\end{figure}

Then we consider a simple stochastic environment whose duration-time
distribution is described by $T=\sum_{i=1}^n T_i$, where $T_i$ obeys
an exponential distribution with mean value ${1}/{n}$. So the mean
value of $T$ is 1 and its variance ${1}/{n}$. When $n=1$, $T$ itself
obeys an exponential distribution, and when $n\rightarrow \infty$,
it returns to a deterministic periodic environment state. A general
form of $f_{\tt f}(t)$ is shown in figure 1(b). Our results show,
when $n=1$ and $k_{\tt u}>k_{\tt f}$, time delays increase the
fitness without reaching a maximum (figure 3(b)). Similarly, after
converting the two switching rates to set $k_{\tt u}<k_{\tt f}$,
$\langle f_{\tt f}\rangle$ monotonically decreases asymptotically to
a value with increasing time-delay. However, in contrast to the
former results, $\langle f_{\tt f}\rangle$ for $k_{\tt u}<k_{\tt f}$
is always larger than that while $k_{\tt u}>k_{\tt f}$. This
phenomenon shows that in this stochastic environment the response
time-delay is always a defect for the fitness, and tuning $k_{\tt
u}<k_{\tt f}$ is simply a better strategy~\cite{Dekel}. When in
cases of larger $n$, results gradually return to the former ones.
Results in case of $n=10$ are shown in figure 3(c), which shows
almost all the characteristics we discuss above. These results show
that the variance may have significant influence on the cells'
fitness just as the averaged value of the fluctuating environment
description. Large variance implies an existence of very short- or
very long-duration environment states. Short ones are filtered by
the cells' phenotype switching system as we discussed above, and
this effect is similar with the static transition. For the long
ones, there are enough time for cells to alter their phenotype with
the prescribed switching rates according to the environment change.
Therefore, these two effects erase maximum or minimum $\langle
f_{\tt f}\rangle$ induced by time delay, and $\langle f_{\tt
f}\rangle$ monotonously converge to the static value with increasing
$\tau$. Similar results hold in non-symmetric periodic environment
description. If the ratio of the duration time of the two distinct
environment states becomes larger, $\langle f_{\tt f}\rangle$ of the
$k_{\tt u}<k_{\tt f}$ also become larger and those of reversed
parameters become smaller accordingly. Finally they separate and
don't intersect just as what happens in the $n=1$ stochastic
environment (data not shown).

In this work, we use a simple stochastic population model with
time-delayed switching-rates conversion to study the influence of
the response time-delay on the fitness of cell population. We find
that in some cases such as $k_{\tt u}>k_{\tt f}$, time delays can
optimize the population fitness, while in some other cases, delay
time don't prove to be so positive. We also find that in the
stochastically fluctuating environment states, variance may be
another important factor determining whether time delay can be an
advantage for the population fitness or not. However, since we only
considered the exponential distribution of duration time above,
other type of stochastic environment should also be considered so as
to give a better understanding of population fitness in such
environment~\cite{Dekel}. Besides, this model only uses the average
growth-rate to account for population fitness while neglecting other
factors such as energy consumption, thus other environment state
factors can be included to give a more comprehensive understanding
of the issue, and experimental verification of mentioned results is
also needed.

{\noindent This work was supported in part by Tsinghua Basic
Research Foundation and by the National Science Foundation of China
under Grant No. 10704045 and No. 10547002.}



\begin{thebibliography}{99}
\bibitem{xue}
We have simulated the effect of distinct response delay-time for two
environment states, while little influence on the trend of the
$\langle f_{\tt f}\rangle$ dependence on $\tau$ is showed, except
the case when the larger $\tau$ dorminates when the system becomes
of a static transition. So here we only consider a specific reponse
delay time.
\end{thebibliography}
\end{document}